\RequirePackage{fix-cm}
\documentclass{article}       
\usepackage{graphicx}
\usepackage{amssymb}
\usepackage{supertabular}
\usepackage{amsmath}
\usepackage{algpseudocode}
\usepackage{algorithm}
\usepackage{tabularx}
\usepackage{multicol}
\usepackage{lineno}
\usepackage{booktabs}
\usepackage{lscape}
\usepackage{natbib}
\usepackage{subfigure}
\usepackage{color}

\begin{document}

\title{A heuristic approach for dividing graphs into bi-connected components with a size constraint
}


\author{Raka Jovanovic      \\
Qatar Environment and Energy Research Institute,\\
 Hamad bin Khalifa University, Doha, Qatar\\
 rjovanovic@qf.org.qa\\
 \\
        Tatsushi Nishi \\
          Graduate School of Engineering Science,\\
           Osaka University,  Osaka,Japan\\
            \\
        Stefan Vo{\ss}\\
        Institute of Information Systems, \\
        University of Hamburg, Germany\\
        and\\
 Escuela de Ingenieria Industrial, \\
Pontificia Universidad Cat\'olica de Valpara\'iso, Chile
        }




\maketitle

\begin{abstract}
In this paper we propose a new problem of finding the maximal bi-connected partitioning of a graph with a size constraint (MBCPG-SC).  With the goal of finding approximate solutions for the MBCPG-SC, a heuristic method is developed based on the open ear decomposition of graphs. Its essential part is an adaptation of the breadth first search which makes it possible to grow bi-connected subgraphs. The proposed randomized algorithm consists of  growing several subgraphs in parallel. The quality of solutions generated in this way is further improved using a local search which exploits neighboring relations between the subgraphs. In order to evaluate the performance of the method, an algorithm for generating pseudo-random unit disc graphs with known optimal solutions is created. The conducted computational experiments show that the proposed  method  frequently manages to find optimal solutions and has an average error of  only a few percent to known optimal solutions.   Further, it manages to find high quality approximate solutions for graphs having up to 10.000 nodes in reasonable time.
\end{abstract}

\section{Introduction}

The bi-connectivity of graphs is essential for many real world applications ranging from power distribution systems, communication networks and many others. The reason for this is that such typologies provide a certain level of resistance to failures, and as a result more robust and reliable systems \citep{Zhang2009812,Moraes20133188,Goldschmidt199697,hu2010generalized}. These benefits are a direct consequence of the fact that such graphs have two disjoint paths connecting any two nodes in the graphs. This property is often called 2-connectivity of graphs. In literature there is differentiation between vertex  and edge 2-connected graphs, where the two paths are vertex  or edge disjoint, respectively. In practice this means that the vertex/edge 2-connected graph stays connected if any single vertex/edge is removed from it. Vertex 2-connectivity is a stronger property in the sense that every such graph is also 2-edge connected but the reverse is not necessarily true. The term bi-connected graph is used for 2-vertex connected graphs. 

Testing if a graph is bi-connected can be done efficiently using the standard algorithm for finding articulation points of a graph based on depth first search  in linear time \citep{hopcroft1973algorithm}. A similar approach has also been used for finding a 3-connected partitioning of  a graph \citep{TriConnected}. Other examples of algorithms for testing bi-connectivity of a graph  exploit the fact that bi-connected graphs have an ear  \citep{EarDecomposition} or  chain  \citep{schmidt2013simple} decomposition.  
There are several types of mixed integer programs based on multi-commodity flow constraints that are used for exploring bi-connectivity, but such models generally have a large number of variables \citep{MorganG08}. There are alternative formulations containing a lower number of variables but having an exponential number of constraints \citep{doForte2013415,buchanan2015integer}.

For many graph optimization problems, like the weighted vertex cover and the independent set, it is sufficient to solve the problem separately on each of the bi-connected components \citep{HOCHBAUM1993203}.  The most commonly used method for partitioning graphs into bi-connected components is Tarjan's algorithm, which accomplishes this task  in linear time \citep{tarjan1972depth}. There are several interesting variations of the original algorithm with similar computational times \citep{Pearce201647}. The problem with these types of algorithms is that it is hard to modify them to a setting where the subgraphs need to satisfy some additional constraints.
  The proposed work is focused on developing a method that can partition a graph into bi-connected subgraphs with a maximal allowed size, but the general concept can be adapted to other interesting constraints.   In current literature there are many practical problems which are modeled using the problem of partitioning graphs into connected subgraphs. 
Some examples are applications in surveillance systems \citep{Borra2015227}, data clustering \citep{shafique2004partitioning} and education \citep{matic2012maximally}. An interesting group of applications comes from the satisfactory graph partitioning problem, like finding communities within social or biological networks,  defense alliances, artificial intelligence development, etc. \citep{Bazgan2010271}. Partitioning problems defined on supply/demand graphs have proven to be essential in modeling systems of interconnected microgrids   \citep{SelAdeq,SelHea,BalancedGSD,SupDem,MIPGSD,Parametric,Tree4}. Clustering in 
   For many of them it would be reasonable to substitute the constraint of connectivity with bi-connectivity, producing the benefit of higher reliability of the system.  This type of extension has frequently been applied  to modeling real-world systems based on general  graph problems  including connectivity. Some examples are power optimization in ad hoc wireless  networks \citep{Moraes20133188} and facility layout problems \citep{Goldschmidt199697}.  A common approach for solving such problems is to start from an approximate solution that is only connected  and extend it with additional  nodes to achieve bi-connectivity, like in the case of the problem of constructing a 2-connected virtual backbone in wireless networks     \citep{BackBone}. 

In case of graph problems containing connectivity constraints,  a standard approach is to grow a partial solution by adding neighboring nodes. One example is the method used for finding the minimal connected dominating set of a graph \citep{Raka2}. The concept  of growth has also been extended to problems where a graph is divided into connected components. In case of such problems the general approach is to grow several subgraphs in parallel  with the constraint that no vertex can be added to more than one of them. The effectiveness of this type of method  is  well presented on the problem of partitioning  supply/demand graphs into connected subgraphs \citep{BalancedGSD,MultiHeuristicGSD,Jovanovic2016317}. It is important to note that the concept of growing a solution gives a high level of flexibility of the method, in the sense of potential applications. Another advantage of growth based algorithms is that they can easily be improved by their extension to metaheuristics like  the ant colony optimization      \citep{dorigo2005ant}, the GRASP algorithm \citep{GRASP} and the variable neighborhood search \citep{hansen2010variable}.

In this paper we introduce a new problem of finding the maximal bi-connected partitioning of a graph with a size constraint (MBCPG-SC). For the newly defined problem we show  NP-hardness. Because of this a growth based algorithm is developed for  MBCPG-SC for finding approximate solutions. More precisely, it is solved using a heuristic procedure that exploits the fact that each bi-connected graph has an ear decomposition. The algorithm is based on a breadth first search (BFS) that also tracks additional properties of the nodes in the BFS tree. The additional information makes it possible to have an efficient way to "grow" a bi-connected subgraph by expanding it with suitable ears. The concept of growing a bi-connected graph using an ear decomposition is also used in case of constructing a fault-tolerant connected set cover problem \citep{Zhang2009812}. In this algorithm the best open ears for extending the current solution are found using the idea of shortest cycles in the original graph. A similar approach is used for the  minimum 2-connected r-hop dominating set problem  \citep{li2010two}.  The proposed  method consists in  growing several subgraphs in parallel, with some auxiliary corrections applied to  the corresponding BFS trees. To improve the performance of the basic algorithm several methods of randomization are developed. The quality of found solutions is further enhanced using  a local search procedure. 

The practical objective of the proposed graph problem and the corresponding solution method is its application to real word problems in the field of smartgrids, more precisely on the underlying wireless networks. Such systems are well presented using unit disc graphs. The proposed graph problem is closely related to the clustering scheme for hierarchical control of wireless networks \citep{banerjee2001clustering,chang2006cluster}. Because of this,  the focus of the numerical experiments is on unit disc graphs. To be able to evaluate the method, an extensive effort has been dedicated to developing an algorithm for generating problem instances with known optimal solutions. Our computational results  show that the proposed method is able to find optimal solutions for small graphs. In case of large graphs (10.000 nodes) the method manages to find solutions within a few percent of error, in reasonable time. 

The paper is organized as follows. In the second section we give the definition for MBCPG-SC and a proof of NP- hardness. In the following section we present the method for growing a bi-connected subgraph. The third section gives  details of the algorithm for parallel growth of several subgraphs and randomization. The next section describes the proposed local search mechanism. In Section 5, we provide details of the data generation mechanism for problem instances with known optimal solutions. Moreover, we discuss results of our computational experiments and provide some conclusions and ideas for future research.

\section{Maximal bi-connected partitioning of a graph with a size constraint}
\begin{figure}[tcb]
\centering
\includegraphics[width=1\textwidth]{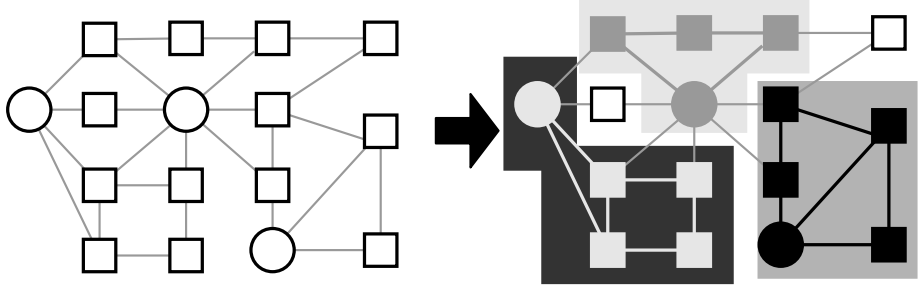}
\caption{Example of a problem instance (left) and solution (right) for the MBCPG-SC with maximal allowed size $M = 5$. The circle nodes represent root nodes.  Different shades of gray  are used for different subgraphs.}
\label{fig:ProblemColor}
\end{figure}

The problem is defined on a graph $G(V,E)$, where $V$ is the set of nodes and $E$ is the set of edges. We also define a set $R \subset V$, whose elements  will be called root nodes. The aim of the problem is to divide the graph $G$ into a set of subgraphs $\Pi = \{\bar{S_1}, \bar{S_2},..., \bar{S_n}\}$, where  $n = |R|$, satisfying the  constraints given in the following text. The notation $S_i$ will be used for the set of nodes that induces subgraph $\bar{S_i}$.  Each of the $S_i$ must contain only one distinct root node $r \in R$. A node $v \in V$ can be an element of at most one $S_i$. The number of nodes in each of the subgraphs $|S_i|$ is less or equal to some constant $M$. The last constraint is that each of the subgraphs $\bar{S_i}$ is bi-connected. The goal of the problem is to find $\Pi$ containing the maximal number of nodes in all the subgraphs.  More formally, we wish to maximize the  sum:
\begin{equation}
\label{ProbSum}
 \sum_{i= 1..n} |S_i|
\end{equation}\\
where each of the subgraphs $S_i$ satisfies 
\begin{eqnarray}
|R \cap S_i | = 1 \\
|S_i| \leq M\\
S_i \cap S_j = \emptyset, i \neq j\\
\bar{S_i} \text{ is bi-connected}.
\end{eqnarray}
In the definition we use notation $|S|$ to indicate the number of nodes in a graph. An illustration of a problem instance and solution for the MBCPG-SC is given in Figure \ref{fig:ProblemColor}. It is important to note that the MBCPG-SC does not produce a partitioning in the strict sense, since some nodes may not be included in any $S_i$.

The MBCPG-SC is a hard optimization problem in the sense that it is NP hard. We give a proof by reduction to a restriction of the problem of maximal partition of supply/demand graphs (MPGSD) \citep{Ito2008627}.  It has been shown that the MPGSD is NP-hard  in the case of a graph containing  only one supply node and having a star structure \citep{Ito2008627}.
For a more comprehensive presentation  we first give the definition of the MPGSD for star graphs, which is a simplified version of the original problem. It is defined for an undirected  star graph $G=(V,E)$  with a set of nodes $V$ and a set of edges $E$. The center node of the graph $s$ is called a supply node and it has a corresponding supply value $sup$. All the other nodes  $u\in D = V\setminus  \{s \}$ are  called demand nodes, and they have a corresponding  demand value  which is a positive integer  $dem(v)$.   The aim is to find a set of nodes  $S \subset D$ which satisfies the constraint that  the supply $sup$ must be greater or equal  to the total demand of nodes in $S$. The goal is to maximize the fulfillment of demands.

\begin{figure}[tcb]
\centering
\includegraphics[width=1\textwidth]{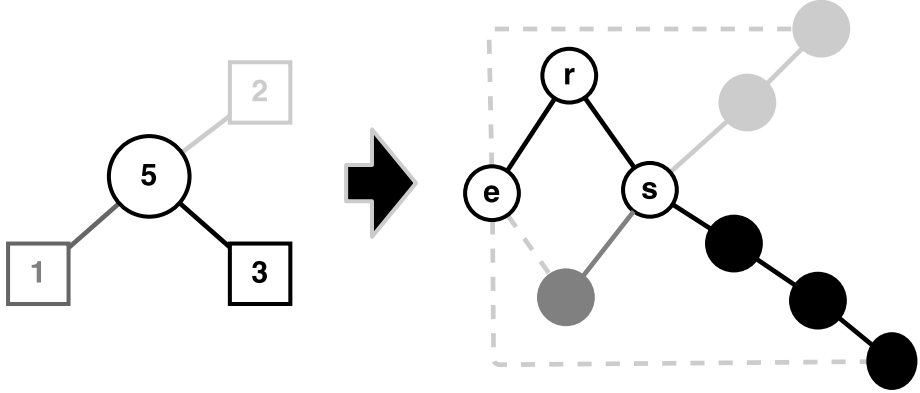}
\caption{Illustration of a conversion of a MPGSD for star graph (left) to MBCPG-SC (right). In the graphic representation for the MPGSD the circle node indicates the supply node and the corresponding supply value. The square nodes indicate the demand nodes and corresponding demand values. Different shades of gray are used to show the changes from nodes (MPGSD) to paths in the MBCPG-SC. In case of the illustration for MBCPG-SC $r$ represents the root node, $s$ the supply node from MPGSD and $e$ the auxiliary node. Dashed lines are used for the auxiliary edges. The size limit MBCPG-SC is $M = 8$.}
\label{fig:ConvertNP}
\end{figure}

In the following text we prove that MBCPG-SC is NP hard by reducing the MPGSD for star graphs to it. Let MPGSD be defined  for a star graph $G'$ having a central supply node $s$ with a supply value $sup$ connected to $n$ demand nodes $d_i$ having demand values $dem_i$. Let us convert this problem to MBCPG-SC in the following way. The parameter  $M$ of MBCPG-SC will be set to $sup+3$ and there will be only one root node $r$. Let us convert the star graph $G'$ to  $\hat{G}(\hat{V}, \hat{E})$ used in MBCPG-SC. First, we include the root node $r$ in $\hat{V}$. The node $r$ is connected to two nodes $s$ (corresponding to the supply node in $G'$) and node $e$ which will be used for some extra edges.  Let us remember that the edge set $E'$ of $G'$ consists of edges $(s,d_i)$. For each demand node $d_i$ we will add a path $(s, n_{i,1}, \cdots , n_{i,dem_i})$. Next, we add an additional edge $(e,d_{i,dem_i})$  for each demand node $d_i$. An illustration of this conversion is given in Figure \ref{fig:ConvertNP}.

Let us make a few observations on the solution of a MBCPG-SC on the convert graph $G'$. First, any solution must contain nodes $e$, $s$ since there must be at least two disjoint paths from $r$ to any other node. Second if any node $n_{i,j}$ is included in the solution $\hat{S}$ all the nodes $n_{i,1}, \cdots , n_{i,dem_i}$ must be included in it. Let us assume the opposite that node $n_{i,k}$ is not included in the solution and that  the nodes $n_{i,j} \in \hat{S}$ for $j\neq k$ are. In such a case the removal of node $e$ from  $\hat{S}$ results in all the nodes $n_{i,l}$ where $l>k$  becoming disconnected from  the graph induced by $\hat{S}$. In the same way, the removal of node $s$ from  $\hat{S}$ results in all the nodes $n_{i,l}$, where $l<k$, becoming disconnected from the graph induced by $\hat{S}$.  As a consequence any $n_{ik} \in\hat{S}$ is connected to $r$ by two disjoint paths (containing only elements in $\hat{S}$)  $(n_{i,k-1}$, $ n_{i,k-2}$ , $\cdots$, $s,r)$ and by path $(n_{i,k+1}$,$n_{i,k+2}$, $\cdots$, $e,r)$. Note that nodes $r$, $s$, $n_{i,1}$,...,  $n_{i,dem_i}$, $e$ form a cycle and as a consequence two disjoint paths exist among any 2 of them.

We will prove that MPGSD on $G'$ can be reduced to MBCPG-SC on $\hat{G}$ by showing that any solution $S'$ of MPGSD has a corresponding solution $\hat{S}$ of the MBCPG-SC such that $\sum_{u \in S'}
 dem(u) = |\hat{S}|-3$ and and vice versa. Let us first prove  the direction MPGSD to MBCPG-SC, by constructing the solution $\hat{S}$ from $S'$.  $\hat{S}$ will consist of nodes $r$, $e$, $s$  and all nodes $n_{u,k}$ were $u \in S'$ and $k=1,.., dem_u$. First, by construction we have  $\sum_{u \in S'} dem(u) = |\hat{S}|-3$.  The only additional case that needs to be considered to prove bi-connectivity of $\hat{S}$  is for two nodes $n_{u,i}$ $n_{v,j}$ such that $u \neq v$. To be more precise we need to prove that there are two disjoint paths connecting them.   From the previous observation we have that each of these two nodes is connected to nodes $e$, $s$  by two disjoint paths. So, nodes $n_{u,i}$ $n_{v,j}$ are connected with disjoint paths $n_{u,i}-s-n_{v,j}$ and $n_{u,i}-e-n_{v,j}$. 

The construction of the solution $S'$ of MPGSD from a solution of $\hat{S}$ of the MBCPG-SC is trivial and is based on the construction of graph $\hat{G}$ and the fact, shown in the previously made observation,  that any solution  $\hat{S}$ consists of $r$, $e$, $s$  and all nodes $n_u,k$  where $k = 1..dem_u$ for some nodes $u \in S' \subset D$.

The main motivation for defining the MBCPG-SC is its application to systems of interconnected microgrids  \citep{MicroGrid}. It has been shown that optimization of self-adequacy of individual microgrids in such systems can be well modeled using the  MPGSD. The proposed problem can also be understood as a partitioning of a power supply network in which all supply nodes (elements of $R$) have a value $M$ and demand nodes (elements of $V \setminus R$) have value 1. The problem with using the MPGSD for this type of systems is that it does not address the problem of failure resistance. In MBCPG-SC we exploit the fact that by strengthening the constraint of connectivity to bi-connectivity, compared to MPGSD, we are able to have a model that produces more robust subsystems. 

Another potential application of the proposed problem is on the hierarchical clustering for wireless networks. In the work of   \citep{banerjee2001clustering}, a network is divided into clusters that can be hierarchically controlled. The basic properties of a cluster are that it has a single control node, the  number of nodes it can contain is limited, and the corresponding  subgraph needs to be connected. It is natural to extend this formulation with an additional constraint of bi-connectivity of  subgraphs to enhance reliability.

\section{Growing bi-connected subgraphs}
\subsection{Definition of open ear decomposition}
\begin{figure}[tcb]
\centering
\includegraphics[width=0.55\textwidth]{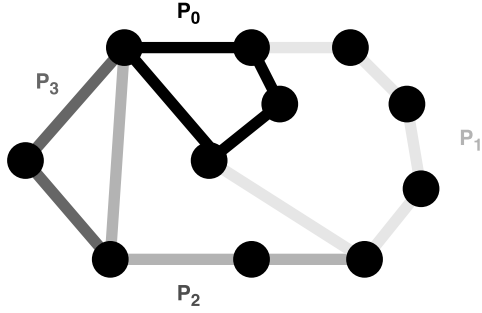}
\caption{Examples of an open ear decomposition for a bi-connected graph. Different shades of gray are used for separate open ears.}
\label{fig:OpenEarDecomposition}
\end{figure}

Since the proposed algorithm is based on the property that a bi-connected graph has an open ear decomposition we start with its definition. An open ear decomposition of a graph $G$ is defined as a series of paths $\bar{P}_0, \bar{P}_i, \dots,\bar{P}_n$ called ears. The term path is used for an ordered sequence ($v_1$, $v_2$, \dots , $v_m$) such that all edges $(v_i,v_{i+1}) \in E$. In the following text we will use the notation $P$ for the set of nodes in $\bar{P}$. The notation $(v^j_1,  \dots,  v^j_{m^j})$ is used for a sequence of nodes in ear $\bar{P}_j$. All the ears $\bar{P}_i$, $i = 1 \dots n$ in the decomposition satisfy  $v^i_j \neq v^i_k$ if $j \neq k$. The exception is the first ear $\bar{P}_0$ which is a cycle $v^0_1 = v^0_n$. For $j>0$, we have that for each of the two terminating nodes $v^j_1, v^j_{m^j} \in P_j$ exists $k,l<j$  such that $v^j_1 \in P_k$, $v^j_{m^j} \in P_l$ where it is not necessarily $l \neq k$. We wish to point out that $v^j_1 \neq v^j_{m^j}$, so each $P_j$ is an open ear. Except for such terminating nodes, there is no $v \in P_i \cap P_j$ if $i \neq j$. Finally, for each $ v \in G$ there exists  at least one $i$ such that $v \in P_i$. If such a decomposition exists for graph $G$ then $G$ is bi-connected. An illustration of an open ear decomposition of a graph is given in Figure \ref{fig:OpenEarDecomposition}. 

\subsection{Algorithm outline}
In the proposed algorithm the idea is to grow a bi-connected subgraph $S$, starting from an initial cycle and extending it with adequate ears.  More formally, we will be iteratively generating a sequence of subgraphs $\bar{S^0} \subset \bar{S^1} \subset \bar{S^2}\subset \cdots$, where $S^{i+1} = S^{i} \cup P^i$ and $P^i$ is an open ear for $S^i$.  It is evident that if a subgraph is generated in this way it will always have an open ear decomposition. Although  it  is possible to develop such an algorithm using a depth first search (DFS) the use of BFS is more suitable since it gives us more control of the size $|P_i|$ of the ear that will be added.  

In the subsequent text we will be using the following notation. 
\begin{itemize}
\item{$N(u)$ the set  of adjacent/neighboring nodes to node $u$ in graph $G$. }
\item{$BFStreeV$/$BFStreeE$ are  the  set of nodes/edges in the BFS tree.  }
\item{$ch(u)$ is the set of children of node $u \in BFStree$ in the BFS tree.}
\item{$par(u)$ is the parent node of $u \in BFStree$ in the BFS tree.}
\item{$desc(u)$ is the set of all descendants of node $u \in BFStree$ in the BFS tree. }
\item{$\bar{p}_b[u,v]$ is defined for nodes $u, v \in BFStree$. It represents the path $(u$, $w_1$,  $w_2$, $\cdots$, $w_n$, $v)$ connecting $u$, $v$  such that  all $w_i \in BFStree$. Brackets "$($"/"$[$"  are used to indicate if $u$, $v$ are included/excluded in the path. The notation $p_b[u,v]$ is  used for the corresponding set of nodes. }
\item{$root(u)$}, is node $v$ which is the first ancestor of $u$, in the BFS tree, such that $v \in S$. In case  $u \in S$, then $root(u) = u$. 
\item{$root(u,P)$}, is node $v$ which is the first ancestor of $u$, in the BFS tree, such that $v \in P$. In case  $u \in P$, then $root(u) = u$. 
\item{$d(u)$} is the length of the path (number of nodes), $p_b[u, root(u))$. Note, that it does not include node $root(u)$. In case $u = root(u)$, we have $d(u) =0$. 
\end{itemize}
It is well known that BFS can be used to find cycles which we exploit in the proposed method. Let us assume that we start the BFS from some initial node $r$.  As we expand the BFS tree, or in other words, new nodes are visited,  the first time we encounter a back-edge $(u,v) \in E \setminus BFStreeE$ an initial cycle $S$ is found. More precisely,  $S$ is acquired by connecting three segments:  path from $r$ to $u$, the back edge $u,v$ and the path from $v$ to $r$. In the proposed notation $S=(p_b[r, u], p_b[v, r))$ and it has all its nodes in the BFS tree.  

In a similar way we can find new open ears. Let us assume the BFS tree is further expanded and a new back-edge $(s,t) \in E \setminus BFStreeE$ has been found, and that at least one of the nodes $s,t$ is not in $S$. It is obvious  that if the following is satisfied, 
\begin{equation}
	\label{ConDifRoot}
	root(s) \neq root(t) 
\end{equation}
then the sequence 
\begin{equation}
\label{PNormal}
\bar{P}(s,t) =  (\bar{p}_b[root(s),t], \bar{p}_b[s, root(t)])
\end{equation}
 will produce a new open ear connected to $S$. As a consequence $S \cup P(u,v)$ will also be a bi-connected subgraph. We will use the notation $P(u,v)$ for the set of nodes corresponding to $\bar{P}(u,v)$. The same procedure can be used to further expand $S$ with new open ears.
 
 The second constraint that exists in the MBCPG-SC is that $|S|<M$. While growing $S$, it can easily be maintained if we only allow adding  ear $P(u,v)$ if the following equation is satisfied 
\begin{equation}
\label{ConLength}
|S| + d(u)+ d(v) \leq M  
\end{equation}
If we adapt the BFS search in a way to always explore node $u$ having the lowest value of $d(u)$ the length $|P|$ of the newly found open ear will be among the shorter ones (except in rare cases).  This is due to the method of construction in which newly  found ears will always have one terminating node equivalent to the currently visited node in the BFS. If $d(M)$ has the largest value of all nodes in the BFS tree which are not in $S$ then the maximal   difference between $|P|$ and the shortest ear will be $d(M)$.  In case of adapting the BFS in this way some additional work will be necessary to update the values of $d(u)$ and $root(u) $ as new ears are added to $S$. It is important to note that after such updates it is possible that several back edges $(u,v)$ that have been previously tested may now satisfy the constraints given in Eqs. \eqref{ConDifRoot},\eqref{ConLength}.  

\subsection{Algorithm}

The algorithm for growing a bi-connected subgraph  based of the idea presented in the previous section will start the BFS from some root node $r$. As the first ear in the decomposition $P_0$ differs from the rest as it is a cycle, some special initialization needs to be done for $r$ and its neighbors $N(r)$. For all these nodes $u$ we will initially set $root(u)=u$. In the adaptation of BFS for growing a bi-connected graph the distance in the BFS tree will have a different meaning. Instead of following the distance of a node $u$ from the root node $r$ we will track the distance from $u$ to the already generated bi-connected subgraph $\bar{S}$. At the initial step we will consider $S = \{r\}$. Let us assume that we have found two nodes $u$, $v$ that satisfy the constraints given in Eqs. \eqref{ConDifRoot}, \eqref{ConLength}, then the first ear $P_0$  can be constructed as 
\begin{equation} 
\label{PRoot}
P_0 = P(u,v) \cup \{r \} 
\end{equation} 

As previously stated in case a new ear $P$ is added to   $S$ the values of $root(u)$, $d(u)$ will need to be updated for some elements of the BFS tree.  It will be necessary to update these values for all $u \in P \cup desc(P)$ in the following way.

\begin{equation}
root(v) =   root(v,P).
\end{equation}

\begin{eqnarray}
d'(v) =   \left\{ \begin{array}{l}
 d'(v)- d'(root(v, P))\,\,\,\,\,\,\,\,\,\,\,\,\,\,\,\,\,\,\, v \notin P\\
 0\,\,\,\,\,\,\,\,\,\,\,\,\,\,\,\,\,\,\,\,\,\,\,\,\,\,\,\,\,\,\,\,\,\,\,\,\,\,\,\,\,\,\,\,\,\,\,\,\,\,\,\,\,\,\,\,\,\,\,\,\,\,\,\,\,\,\,\,\,\,\, v \in P\\
 \end{array} \right.
\end{eqnarray}
It is important to note that the proposed correction for functions $d(u)$ will produce approximations to the exact 
distance $d'(u)$. For the function $d'$ we have that  $d'(u)\geq d(u)$ since it is possible to have an alternative path to $S$ using some back edges which is shorter. A consequence of this is that the constraint defined in  Eq.\eqref{ConLength} will never give false positives if function $d'$ is used instead of $d$. By using this approximate approach it is possible to have a simpler and less computationally expensive implementation. 

In the standard BFS there is no change in the distance for visited nodes and no node is re-visited. In the proposed adaptation of BFS such changes can occur and some revisits are necessary. It is possible to use a heap or similar structure instead of a queue to always test the node $u$ with the lowest value $d'(u)$. In practice this is not necessary especially since we are only using an approximation to $d(u)$. On the other hand the need for retesting some nodes further increases the complexity of implementation. Both of these issues are addressed simultaneously using the following approach. First, nodes will be re-added to the queue as a new ear is added to $S$ and their re-evaluation is needed. Since it is possible for the same node to be added multiple times to the queue due to the addition of multiple ears, an additional value will be used to track if an evaluation is needed. The algorithm for growing a bi-connected subgraph is better understood by observing Algorithm \ref{Alg:Grow}.
\begin{algorithm}
\begin{algorithmic}
\Procedure{BFSGrowBiConnected} {G, r}
\State{For all $u \in G$  initialize $Dist, Eval, Parent$ }
\State{Initialize all $u \in r \cup N(R)$ }
\State{Add all $N(R)$ to $Q$ }
\While{$Q$ is not empty}

 \State{$current = Q.dequeue()$} \Comment{Using Queue (FIFO) structure}
 \If{$current.Eval \wedge (M-|S| \leq current.Dist)$}
 \ForAll{$u \in N(current)$}
 \If{$(u,current)$ is BackEdge)}
 	\If {$u,v$ produce an open ear satisfying Eq. \eqref{ConDifRoot}, \eqref{ConLength}}
	 	\State{Set $P(u,c)$ based on Eq. \eqref{PNormal} or  Eq.\eqref{PRoot}}
 		\State{$S = S \cup P(u,c)$} 
	 	\State{$Update(P,Q)$} 
	 	\State{Exit procedure if $S = M$}
 	\EndIf
 \Else
 	\State{$u.[Root, Dist]= [current.Root, current.Dist+1]$} 
 	\State{Update parent, child relations for $u$, $current$ } 
	\State{$Q.enqueue(u)$ } 
 \EndIf
 \EndFor
  \State{$current.Eval = false$}
  \EndIf
\EndWhile
\EndProcedure
\end{algorithmic}
\caption{\label{Alg:Grow} Pseudo code for growing a bi-connected subgraph}
\end{algorithm}

The proposed algorithm starts with a standard BFS initialization of the distance, parents and descendants for all the nodes with the additional property of the need for evaluation. Initially all the values of $Eval$ will be set to $true$. An auxiliary structure is used to store all the properties of individual nodes, which can be accessed and updated using  the node id. Next, we initialize the root $r$ and all its neighbors $N(r)$ as previously described and all nodes in $N(r)$ are added to the queue $Q$. The main loop is executed for each node $current$ in $Q$ until $Q = \emptyset$.  For each such node we first check if an evaluation is needed and if so all its neighbors $N(current)$ are evaluated. For each $u \in N(current)$ we check if $(current, u)$ is a back-edge. In case it is not we add $u$ to the $Q$ as in the BFS, and we set $root(u) = root(current)$. In case $u$ is a back-edge we check if  $P(u,current)$ is an open ear connected to $S$. If this is true the subgraph $S$ is extended with $P(u, current)$ and necessary updates are performed using procedure  $Update(P,Q)$. After all the elements of $N(current)$ are visited the evaluation of node $current$ is complete and we set $current.Eval = false$.  

The update procedure for a newly found open ear, $Update(P,Q)$,  is used to change the state of $Q$ and nodes based on the $P(u, current)$.  The details of the procedure are given in Algorithm \ref{Alg:UpdateAddEar}.
\begin{algorithm}
\begin{algorithmic}
\Procedure {Update }{P,Q} 
\ForAll{($u \in P $)}
			\State{$u.[Root, Dist] = [u, 0]$}
 		    \State{$UpdateBFSBranch(u,u,0)$}
 		    \State{$Q.Enqueue(u)$}
\EndFor
\EndProcedure

\Procedure {UpdateBFSBranch}{ u, root, dist} 
\ForAll{($v \in ch(u) $)}
		\If{$v \notin S \wedge v \in BFStree$ } 
			\State{$u.[Root, Dist, Eval] = [root, dist+1, true]$}
 		    \State{$UpdateBFSBranch(v,root, Dist+1)$}
  		    \State{$Q.Enque(u)$}
  		 \EndIf   
\EndFor
\EndProcedure
\end{algorithmic}
\caption{\label{Alg:UpdateAddEar} Update procedure for adding an ear to subgraph $S$.}
\end{algorithm}
In it, for all $u \in P$ the distance is set to $d.Dist = 0$. Each node $u$ now becomes a root of a new potential ear, so we set $u.root = u$. For each node $u \in P$ we wish to update the branch of the BFS tree whose root is $u$. 

 This is done using a recursive procedure $UpdateBFSBranch(u, root, dist)$.  In it we go through all the BFS descendants $v$ of $u$ that are not already in $S$ and set the $v.root = root$ and $v.dist = dist+1$. By doing so the node properties are set to the values that correspond to the new state of $S$. For each such node re-eavaluation is needed to check if new open ears have been created so we set $v.Eval = true$. The addition  to the queue $Q$ is done after the recursive call $UpdateBFSBranch(v, root, dist+1)$  for updating the descendants. This order is important since nodes will be added to the queue in reverse order of their distance and as a consequence shorter potential ears are checked first. By doing so $S$ is more gradually grown.

\section{Algorithm for MBCPG-SC}

When solving the MBCPG-SC multiple subgraphs $\bar{S_1}, .., \bar{S_n}$ should be grown together. The idea of the algorithm is to randomize this process. The randomization is done on two levels. First,  the growth of individual subgraphs $\bar{S_i}$ should be randomized. This can simply be done by adding an additional parameter $p_0 \in [0,1]$  and a corresponding random variable $p\in [0,1]$ which will be used to decide if a valid open ear is added to $S$ or not. It is important to note that by not adding an open ear $P_k$ at iteration $k$  does not necessarily exclude the nodes inside $P_k$ from the corresponding subgraph $\bar{S}$. This is due to the fact that they can be a part of some ear $P_l$ that will be added to $S$ at a later iteration. 

It is evident that the growth of subgraphs $\bar{S_1}, \cdots , \bar{S_n}$ is interdependent since a  node $u$ can only be an element of a unique $S_i$. On the other hand, the growth of some $S_i$ will also effect the direction of expansion of the BFS tree of neighboring subgraphs. Because of this the second type of randomization should effect the speed and order in which all the $S_i$ will be grown. The proposed method can be better understood through the pseudocode given in Algorithm \ref{Alg:GenerateSolution}.
\begin{algorithm}
\begin{algorithmic}
\Procedure {GenerateMBCPG-SC}{roots} 
\State{For all $i$ Set $root^j_i.Unavalable = true$ for all $S_j$ where $i\neq j$}
\State{Initialize all $S_i$ for $roots_i$}
\While{Can Expend Some $S_i$}
\State{Select Random Expansion Length $MaxL \in (2, MaxExpLength)$}
\State{Select Random $S_i$ from Expandable}
\State{$cLength = 0$}
\Repeat
\State{$P = BFSGrowSingleEarBiConnected(S_i, G)$}
\State{Apply $UpdateBFSTree(S_j,P)$ for all $j\neq i$ } 
\State{$cLength =  cLength+ |P|$}
\Until{$(cLenght \geq MaxL)$ $\wedge$ $CanExpand(S_i)$) }
\EndWhile
\EndProcedure
\end{algorithmic}
\caption{\label{Alg:GenerateSolution} Randomized method for generating a solution for MBCPG-SC $S$.}
\end{algorithm}

In the proposed algorithm  separate BFS trees and corresponding auxiliary structures exist for each of the $S_i$. The auxiliary structure for tracking node properties is extended with a property $Available$ for nodes, which is used to indicate if a node has been added to some of the other subgraphs. Initially this value is set to $true$ for all nodes in all subgraphs. The first step, is setting $root^j_i.Available = false$ for all the $S_j$ where $i \neq j$. Here we use the notation $u^j_i$ for node $u_i$ in the auxiliary structure of subgraph $\bar{S_j}$. Next, the $BFStree$ is initialized for each subgraph $\bar{S_i}$ using the corresponding $root_i$ as described in the previous section. The main loop is repeated until no subgraph can be further expanded. At each iteration a random subgraph $\bar{S_i}$ is selected for expansion of a maximal allowed size $MaxL$. $MaxL$ is a random variable from some range $[2,MaxExpLength]$. The goal of the inner loop is to extend $S_i$ with at least $MaxL$ nodes, or until it is not possible to extend it. In this loop an adaptation of the procedure presented in the previous section  is used in the form of function $BFSGrowSingleEarBiConnected(S_i,G)$ which only adds a single ear $P$ to $S_i$. This function does not consider nodes having $Available = false$. After an ear $P$ is added to $S_i$, the  BFS tree and corresponding structures need to be updated for all $S_j$ where $i \neq j$ using procedure $UpdateBFSTree(S,P)$. 

The update procedure needs to perform several tasks. First, all the nodes $u \in P$ need to be made unavailable for the growth of $S_j$, where $j \neq i$, which can be done by setting $u.Available = false$ in the corresponding auxiliary structures. Secondly, all the nodes in $v \in desc(u)$ which have been cutoff from the BFS tree (for some $S_j$, $i \neq j$), by the removal of $u$, need to be re-initialized so they can potentially be re-added to the BFS tree. This is done by setting $v.Distance = INF$ and $v.Eval = true$. Although there is a potential that a node $v \in desc(u)$  may be reached by continuing the growth of the BFS tree, there is no guarantee for this.  The question is which nodes need to be re-evaluated to make this possible. It is obvious that if a back-edge  $(w, v)$ existed for the BFS tree that node $w$ can be reconnected to $v$. On the other hand if such a back edge existed and $root(w) != root(v)$ then the corresponding open ear would have been already added to $S$. The only potentially disregarded back edges  are of the type $root(w) = root(v)$.  Because of this  the nodes in $a \in p_b(u,root(u)) \setminus u$  should be re-evaluated if  $a.Eval = false$, since they can  establish a connection with $v$. The details of the update procedure for deleting a node are given in Algorithm \ref{Alg:DeleteNode}.

\begin{algorithm}
\begin{algorithmic}
\Procedure {UpdateBFSTree}{$j$, $P$} 
\ForAll{ $u \in P$ }
\State{ $u.Available = False$}
\State{Update $Parent$, $Descent$ relation for $u$}
\ForAll{$c \in p_b(u, root(u))\setminus \{u\}$}
   \If{$c.Eval = false$}
	\State{$c.Eval = true$}	
	\State{$Q.enqueue(c)$}
\EndIf
\EndFor

\ForAll{$v \in desc(u)$}

\State{ $v.[Eval, Dist] = [false, INF]$}
\State{Clear parent, child relations}
\EndFor
\EndFor
\EndProcedure

\end{algorithmic}

\caption{\label{Alg:DeleteNode} Randomized method for generating a solution for MBCPG-SC.}
\end{algorithm}

\section{Local Search for MBCPG-SC}

The algorithm in the previous section gives us a method for generating a single solution for MBCPG-SC.  A simple way of finding higher quality solutions is to perform multiple runs with different random seeds and selecting the best one. The problem with this approach is that no experience is gained from previously generated solutions and as a consequence a very high number of them needs to be generated to get good quality ones. An alternative approach is to develop a local search procedure which improves sections of already generated solutions. The basic idea of the proposed local search is to regrow only a subsection $I \subset \Pi$ of the previously best found solution. Ideally, we wish to do this only  for  subsections for which an improvement is possible. The growth procedure presented in the previous section can easily be adapted to such a setting. 

Let us define the set on non-located nodes as
\begin{equation} 
NonLoc = V  \setminus \bigcup\limits_{i=1..n} S_i 
\end{equation} 
It is obvious that $I$ should contain at least one $\bar{S_i}$ such that $|S_i| < M$. The other requirement is that $I$ has a potential to expand to some $u \in NonLoc$. For simplicity let us assume that there is only one non-located node $u \in NonLoc$, and there is only one $\bar{S_i} \in I$ such that $|S_i|<M$. Since $\bar{S_i}$ must be bi-connected there must be a $path(root_i,u)$ having only nodes in $I$. We can define the set of neighboring nodes of a subgraph in the following way
\begin{equation} 
N_s(\bar{S}) = \{u \mid (u \in V)\wedge (u \notin S) \wedge (\exists(v \in S)((u,v) \in E) \} 
\end{equation} 
It is evident that for an appropriate $path(root_i, u)$ to exist, $u$ must satisfy $u \in N_s(S_j)$ for some $\bar{S_j} \in I$. Using this logic, we can specify $I$ containing $m$ subgraphs based on this necessary condition. Select one $\bar{S_i} \in \Pi$ having $|S_i|<M$ and $\bar{S_j} \in \Pi$ such that it  $N_s(\bar{S_j}) \cap NonLoc \neq \emptyset $. Select $m - 2$ random subgraphs from $\Pi \setminus \{ \bar{S_i}, \bar{S_j} \}$. We will call this method for selecting the elements of $I$ growth random ($GrowR$).

This method for specifying $I$ satisfies the necessary condition but further constraints can be added to make the local search more efficient. Let us observe a graph $G'(V',E')$ induced by $\Pi$ in the following way.  Each node in $V'$  corresponds to a subgraph $\bar{S_i} \in \Pi$. An edge $(\bar{S_i},\bar{S_j})$ is in $E'$  if there exist nodes $u \in S_i$, $v \in S_j$ such that $(u,v) \in E$. If we are re-growing only two subgraphs $\bar{S_i}, \bar{S_j}$ there is a potential that nodes may be exchanged between  them  if $(\bar{S_i}, \bar{S_j}) \in E'$, and more generally $S_i, S_j$ will influence each other's growth. Transitively, the influence will extend to all subgraphs in a connected subgraph $S' \subset G'$.

\begin{figure}[tcb]
\centering
\includegraphics[width=0.85\textwidth]{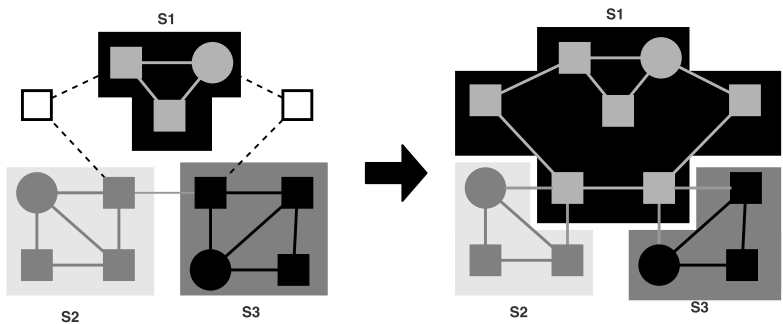}
\caption{An illustration of an exchange between subgraphs through non-located nodes. Different shades of gray are used to indicate different subgraphs. Dashed lines represent non-located paths (left side). The edge sets used for defining  $G'$ and $\hat{G}$  are $E' = \{(S_2, S_3)\}$ and $\tilde{E} = \{(S_1, S_2), (S_1, S_3)\}$, respectively. }
\label{fig:NonLocPath}
\end{figure} 

An exchange can occur between subgraphs $\bar{S_i}, \bar{S_j}$ in the same setting even if $(\bar{S_i},\bar{S_j}) \notin E'$  through nodes in $NonLoc$.     Let us assume that for some $u \in S_i$, $v \in S_j$ there is a connecting $path(u,v)$ which only contains nodes in $NonLoc$. In such a case there is a chance that node $v$ can be added to $S_i$ using this path. We give an illustration   in Figure \ref{fig:NonLocPath}. Let us define $\tilde{E}$ as a set of edges, where  $(\bar{S_i},\bar{S_j}) \in \tilde{E}$ only if a connecting path exists containing only  non-located nodes.  Let us now define a new graph $\hat{G}(V', \hat{E})$ using the following edge set 
\begin{equation}
\hat{E} = E' \cup  \tilde{E}
\end{equation} 
Now, we can say that for the same $I$ an exchange between subgraphs $\bar{S_i}$, $\bar{S_j}$  can only occur if and only if $(\bar{S_i}, \bar{S_j}) \in \hat{E}$ and more generally $S_i, S_j$ will influence each other's growth. Transitively the same applies to any connected subgraph $I \subset \hat{G}$. A consequence of this is that  regrowing should only be considered for a connected (in $\hat{G}$) subsections  $I$, since each isolated section can be treated separately. 

Using graph $\hat{G}$ we can define a method for generating a set $|I| \geq m$ suitable for regrowth in the following way. First select an initial  random node $I = \{\bar{A_0}\}$ such that $|A_0| <M$. Note as a reminder that $\bar{A_0}$ represents a subgraph in $\Pi$. Iteratively, we add a random node $\bar{A_i}$ such that it is connected (in $\hat{G}$) to at least one node  $\bar{A_j} \in I$. When $|I|=m$, we check if at least one node $\bar{A_j} \in I$  satisfies  $N_s(\bar{A_j}) \cap NonLoc \neq \emptyset$. If this is true  $I$ is accepted as the set that will be regrown. In case this is not true a new set $I$ is generated in the same way. If after $N$ attempts no such $I$ is generated $m$ is increased by one. This process is repeated until $I$ satisfying the necessary constraints is generated.  We will call this method for selecting the elements of $I$ growth neighbor ($GrowN$).

The method based on the local search  for the MBCPG-LS is given in Algorithm \ref{Alg:LocalSearch}.
\begin{algorithm}
\begin{algorithmic}
\State{ Generate solution $\Pi$ using $GenerateMBCPG-SC$}
\State{ $BestSolution = \Pi$}
\State{Calculate $\hat{G}(V', \hat{E})$ for $BestSolution$}
\While{Not Termination Condition}
	\State{Randomly select $m$ from $[2, RegrowSize]$ }	
	\State{Randomly select $\bar{S_i}$ such that $|S_i|<M$}
	\State{Generate $I$ based on $S_i$, $m$ and $\hat{G}$}
	\State{$\Pi' = RegrowPartial(I,BestSolution)$}
	\If{$|\Pi'| \geq |BestSolution|$ }
	\State{ $BestSolution = \Pi$}
	\State{Calculate $\hat{G}(V', \hat{E})$ for $BestSolution$}
	\EndIf
\EndWhile
\end{algorithmic}
\caption{\label{Alg:LocalSearch} Local search based algorithm for finding a high quality solutions for MBCPG-SC.}
\end{algorithm}
The methods starts with generating an initial solution using procedure $GenerateMBCPG-SC$ which is set as the $Best Solution$. For this solution we generate the graph $\hat{G}$ as described in the previous text. In the main loop we first select a random subgraph $\bar{S_i}$ such that $|S_i|<M$ and a random number $m \in [2, RegrowSize] $ which indicates how many subgraphs  will be regrown. Next, we generate a random  subgraph  $I \subset \hat{G}$ based on the node $S_i$ containing $m$ nodes  using one of the two methods $GrowR$ or $GrowN$. A new solution $\Pi'$ is acquired by regrowing $I$ using procedure $RegrowPartial(I,BestSolution)$. This procedure is the same as $GenerateMBCPG-SC$ with the following changes. First, for all nodes  $u \notin NonLoc \cup I$ we set $u.Available = false$. Next, only subgraphs in $I$ are grown and considered in the update procedure. The rest of the subgraphs are left unchanged. 
The next step in the main loop consists in  testing if $|\Pi'| \geq |BestSolution|$. In case this is true the best found solution is updated and the graph $\hat{G}$ is recalculated. Note that we have allowed the update of the best solution even if it has the same quality as $\Pi$ to diversify the search. The main loop is repeated until a maximal number of iterations or no further improvement can be achieved.

\section{Computational Experiments}

In this section we present the results of the computational experiments used to evaluate the  performance of the proposed  method. We have compared the basic randomized growth algorithm with its extension using the local search procedure with and without exploiting the subgraph neighborhoods. All the algorithms  have been implemented in C\# using Microsoft Visual Studio 2015. The source code and the execution files have been made available at \citep{Data}. The calculations have been done on a machine with Intel(R) Core(TM) i7-2630 QM CPU \@ 2.00 Ghz, 4GB of DDR3-1333 RAM, running on Microsoft Windows 7 Home Premium 64-bit. 

\subsection{Test instance generation}
In the evaluation of the proposed algorithm our focus is on unit disc graphs due to there close relation with wireless networks. As  stated above the proposed problem can be used to find   clustering schemes for hierarchical control of such systems with enhanced fault tolerance.  
Due to the fact that this is a newly defined problem there are no standard benchmark instances available for comparison it was necessary to also generate  problem instances with known optimal solutions. In the generation procedure our goal is to generate graphs that are "as random as possible". To be more precise we attempt to have the node positions close to the uniform probability distribution and to have the number of edges that are adjacent to each node relatively balanced. To achieve this we use the following procedure. 

Let us say that we are generating a problem instance having $n$ root nodes and $M$ is the maximal allowed number of nodes in a subgraph, having a solution in which $\Pi$ contains all the nodes. 
For problem instances of this form  we generate unit disc graphs inside of a box  $B$ with dimensions $1$, $1$. 
The value for the length for establishing edges is $d = 1/(\sqrt{\alpha n M})$. $\alpha$ is an additional parameter used to vary the density of the graph used in specifying different data sets. The first step is generating  $n$ random bi-connected unit disc graphs having $M$ nodes.  As our aim is to have the nodes of graph $G$ spread over the entire area of the box $B$ having an area equal to $1$. We want to have each of the subgraphs $G_i$ covering an area of close to $1/n$. To achieve this  for each such graph we first define a box with dimensions $1/(\sqrt{n} R)$, $\sqrt{n}R$  where $R$ is randomly selected  from the interval $(0.5, 1)$ used to vary the shape of $G_i$.  For this box a random bi-connected graph $G_i$ is generated. This is done by repeatedly generating $M\cdot \delta$ random points and checking if the resulting unit dist graph contains a bi-connected subgraph $G_i$ containing $M$ nodes. To increase the probability of having nodes of the graph covering an area close to $1/n$, four nodes have been forced to random positions at each edge of the  box.  To achieve this a high number of graphs needs to be generated, especially in the case of a high value of $M$. This is the reason for the additional points, specified using parameter $\delta = 1.1 $, in the box, since it greatly reduces the number of graphs that are generated.  Out of all the nodes in $G_i$ one would be randomly selected as the $root_i$.

The next step in generating a problem instance with a known solution is combining graphs $G_i$. Let us remember that by the method of construction for each subgraph $G_i$ contains $M$ nodes having some values of coordinates $(x, y)$, as a consequence we can easily translate $G_i$ by a vector $(t_x, t_y)$ by changing all node positions to $(x+t_x, y+t_y)$. The basic idea of combining the graphs $G_i$ is to randomly position (translate) them in a way that all the nodes fit in a box having dimensions $1$, $1$. As previously stated our goal is to have a relatively balanced number of edges for all the nodes. Secondly, we want to distribute the nodes over the entire box $B$.  This is achieved through the following iterative procedure. Initially, the graph $G$ is set to  $G_0$. At each iteration a new graph $G_i$ is set at a random position and added to $G$, but some constraints must be satisfied. First, all the points in $G \cup G_i$ must fit into a box having dimensions $1$,$1$. The number of new edges $ne$ connecting $G$ and $G_i$, has to satisfy $ne \geq  MinCon$.  To achieve this multiple random positions  have to be tested for each subgraph $G_i$. The parameter $MinCon$  is used to bound the allowed number of new edges which is related to the number of edges $|E_i|$ in the graph as $MinCom = max(3, \gamma |E_i|)$. Empirical tests have shown that values $\gamma = 0.2$ produce satisfactory results. For each  graph $G_i$ a $1000$ positions are tested and the one satisfying the constraints and having the lowest number of $min_{j= 1..i} |E_j|$ would be chosen for adding to $G$. The last step is randomly enumerating all the nodes in $G$. 

\subsection{Experiments}

To have an extensive evaluation of the proposed algorithm, we have generated a wide range of problem instances for different numbers of root nodes $n$ (5-100) and maximal allowed number of nodes in a subgraph $M$ (5-100). For each pair $n$, $M$ 40 problem instances have been generated using the algorithm presented in the previous subsection with different values for the seed of the random number generator. We have generated two problem sets varying in the level of connectivity by setting $\alpha = 1.5, 2$.  For each of the problem instances a single run of the growth based algorithms is performed. This is done for the two local search methods based on $GrowR$ and $GrowN$ for generating the set of subgraphs $I$ which will be regrown. 

The same set of parameters for specifying the algorithm is used for all the problem sizes. The value of the parameter for accepting an ear is $p_0 = 0.5$ in the growth algorithm.  The parameter specifying the maximal expansion of a subgraph  at one step is $MaxExpLength = 12$.  The value $RegrowSize = 9$ was used for the the upper bound on the value $m$ which is used for generating the size $|I|$ of the set of subgraphs that will be regrown. These values have been chosen empirically after a wide range of values have been tested in computational experiments. In all of the computational experiments the algorithm would execute until  10.000 iterations (generated solutions) have been reached  or  more than 2000 iterations have been completed without an improvement to the best found solution. 

The results of the computational experiments focus on the quality of found solutions and the computational cost which can be seen in Tables \ref{table:Alpha2}, \ref{table:Alpha15}. The evaluation of solution quality is done using the  average normalized error of the found solutions compared to the known optimal ones, for each of the used methods. More precisely, for each of the 40 test instances, for each pair $(n,M)$,  the normalized error is calculated by $(Optimal - found)/Optimal \cdot 100$, and we show the average values.  To have a better comprehension of the performance we have also included the standard deviation, maximal errors and the number of found optimal solutions (hits).  The computational cost is analyzed through  the average execution time  needed to find the best solution. To have a better understanding of these execution times,  we have also included the average number of iterations (number of generated solutions) to find the best solution and the corresponding standard deviation in Tables \ref{table:Alpha2}, \ref{table:Alpha15}.

\begin{table*}[htb]
\footnotesize
\center
\caption{\label{table:Alpha2}
Comparison of the performance of the $GrowR$ and $GrowN$ methods for  unit disc graphs with $\alpha = 2$.}
\begin{tabularx}{430pt}{X*{10}{c}}

\toprule
Roots X M&  \multicolumn{2}{c}{$Avg(Stdev)$}  & \multicolumn{2}{c}{$Max$} & \multicolumn{2}{c}{$Hits$}& \multicolumn{2}{c}{$Avg Iter$}& \multicolumn{2}{c}{$Avg Time(ms)$}\\

& $R$ & $N$  &  $R$ & $N$ &  $R$ & $N$& $R$ & $N$&  $R$ & $N$\\

\midrule
5 X 5 & 0.30(1.38) & 0.30(1.38) & 8.00 & 8.00 & 38 & 38 & 122(47) & 63(41) & 12 & 8 \\
5 X 10 & 0.15(0.69) & 0.00(0.00) & 4.00 & 0.00 & 38 & 40 & 357(224) & 269(228) & 40 & 31 \\
5 X 25 & 0.78(1.32) & 0.96(1.61) & 5.60 & 7.20 & 24 & 23 & 884(435) & 644(170) & 236 & 185 \\
5 X 50 & 1.12(1.46) & 0.90(1.41) & 8.00 & 8.00 & 11 & 14 & 1428(1160) & 1455(364) & 851 & 940 \\
5 X 100 & 1.40(1.39) & 1.38(1.64) & 5.00 & 6.00 & 5 & 7 & 1707(851) & 1738(215) & 2472 & 2547 \\
\midrule
10 X 5 & 0.50(1.40) & 0.25(0.80) & 6.00 & 4.00 & 35 & 36 & 393(903) & 359(507) & 34 & 33 \\
10 X 10 & 0.43(0.80) & 0.28(0.63) & 4.00 & 3.00 & 28 & 32 & 1034(928) & 848(775) & 162 & 162 \\
10 X 25 & 2.78(2.03) & 2.30(2.26) & 9.60 & 9.60 & 1 & 5 & 1810(1031) & 1569(594) & 748 & 680 \\
10 X 50 & 2.04(1.34) & 1.51(1.19) & 6.80 & 5.00 & 0 & 1 & 2775(441) & 2946(1389) & 2497 & 2730 \\
10 X 100 & 1.92(1.30) & 1.70(1.35) & 6.30 & 6.30 & 0 & 0 & 3470(555) & 3479(1997) & 7883 & 7870 \\
\midrule
25 X 5 & 3.36(3.13) & 1.98(2.71) & 9.60 & 8.80 & 12 & 21 & 2113(3057) & 1635(686) & 246 & 207 \\
25 X 10 & 2.19(1.42) & 1.41(1.18) & 5.20 & 4.00 & 2 & 7 & 2268(2228) & 1731(890) & 431 & 399 \\
25 X 25 & 3.35(1.31) & 2.04(0.87) & 6.88 & 4.16 & 0 & 0 & 4475(3739) & 4494(1885) & 2221 & 2248 \\
25 X 50 & 2.91(0.98) & 2.41(0.95) & 4.88 & 4.56 & 0 & 0 & 6225(2062) & 4480(1170) & 6842 & 4985 \\
25 X 100 & 2.59(0.93) & 2.32(1.04) & 6.00 & 5.16 & 0 & 0 & 7220(2221) & 6105(1092) & 22812 & 18533 \\
\midrule
50 X 5 & 6.47(2.08) & 3.08(1.97) & 10.40 & 8.00 & 0 & 2 & 3669(3037) & 4020(1764) & 746 & 728 \\
50 X 10 & 2.93(1.16) & 1.37(0.82) & 6.00 & 3.20 & 0 & 1 & 5385(2962) & 3548(248) & 1681 & 1118 \\
50 X 25 & 3.88(0.86) & 2.50(0.74) & 5.68 & 4.16 & 0 & 0 & 7779(1800) & 7322(1683) & 5911 & 4801 \\
50 X 50 & 3.91(0.76) & 2.79(0.70) & 5.68 & 4.40 & 0 & 0 & 8822(813) & 7966(1921) & 15347 & 12788 \\
50 X 100 & 3.51(0.79) & 2.47(0.59) & 5.26 & 4.40 & 0 & 0 & 9295(228) & 8995(916) & 42597 & 38596 \\
\midrule
100 X 5 & 8.87(1.53) & 4.87(1.19) & 13.00 & 7.40 & 0 & 0 & 5333(2432) & 5829(832) & 2681 & 2256 \\
100 X 10 & 4.16(0.75) & 1.92(0.74) & 6.00 & 3.40 & 0 & 0 & 8100(1171) & 6944(2128) & 6452 & 3875 \\
100 X 25 & 5.51(0.88) & 2.99(0.47) & 8.00 & 4.04 & 0 & 0 & 9449(306) & 9127(475) & 17044 & 11720 \\
100 X 50 & 5.44(0.70) & 3.26(0.44) & 6.94 & 4.82 & 0 & 0 & 9710(240) & 9640(212) & 39697 & 29134 \\
100 X 100 & 5.00(0.82) & 3.15(0.49) & 6.84 & 4.31 & 0 & 0 & 9844(168) & 9731(255) & 100213 & 78981 \\
\bottomrule
\end{tabularx}
\end{table*}

\begin{table*}[htb]
\footnotesize
\center
\caption{\label{table:Alpha15}
Comparison of the performance of the $GrowR$ and $GrowN$ methods for  unit disc graphs with $\alpha = 1.5$.}
\begin{tabularx}{430pt}{X*{10}{c}}

\toprule
Roots X M&  \multicolumn{2}{c}{$Avg(Stdev)$}  & \multicolumn{2}{c}{$Max$} & \multicolumn{2}{c}{$Hits$}& \multicolumn{2}{c}{$Avg Iter$}& \multicolumn{2}{c}{$Avg Time(ms)$}\\
& $R$ & $N$  &  $R$ & $N$ &  $R$ & $N$& $R$ & $N$&  $R$ & $N$\\

\midrule
5 X 5 & 0.10(0.62) & 0.10(0.62) & 4.00 & 4.00 & 39 & 39 & 158(150) & 153(124) & 10 & 10 \\
5 X 10 & 0.00(0.00) & 0.10(0.44) & 0.00 & 2.00 & 40 & 38 & 254(147) & 211(180) & 27 & 28 \\
5 X 25 & 0.38(0.69) & 0.18(0.42) & 2.40 & 1.60 & 29 & 33 & 511(273) & 526(480) & 137 & 153 \\
5 X 50 & 0.05(0.13) & 0.06(0.17) & 0.40 & 0.80 & 35 & 35 & 911(252) & 542(456) & 536 & 341 \\
5 X 100 & 0.17(0.31) & 0.16(0.43) & 1.60 & 2.60 & 23 & 26 & 1305(1029) & 875(502) & 1772 & 1243 \\
\midrule
10 X 5 & 0.75(1.53) & 1.00(2.10) & 6.00 & 8.00 & 31 & 31 & 624(434) & 349(185) & 49 & 35 \\
10 X 10 & 0.65(1.15) & 0.33(0.75) & 5.00 & 4.00 & 26 & 31 & 807(457) & 578(464) & 126 & 121 \\
10 X 25 & 0.70(1.01) & 0.49(0.82) & 4.40 & 3.60 & 18 & 24 & 1548(820) & 1269(324) & 659 & 595 \\
10 X 50 & 0.39(0.55) & 0.38(0.40) & 2.80 & 1.60 & 15 & 13 & 2429(1540) & 1914(1255) & 2421 & 2043 \\
10 X 100 & 0.50(0.75) & 0.38(0.60) & 3.00 & 2.80 & 6 & 13 & 2859(1494) & 2431(1869) & 7150 & 6134 \\
\midrule
25 X 5 & 3.48(1.47) & 2.18(1.51) & 6.40 & 6.40 & 0 & 6 & 1581(450) & 1336(838) & 208 & 182 \\
25 X 10 & 1.17(1.01) & 0.76(0.79) & 4.40 & 2.80 & 3 & 13 & 2256(1616) & 1664(714) & 516 & 496 \\
25 X 25 & 1.18(0.74) & 0.80(0.69) & 3.52 & 2.88 & 0 & 4 & 4392(2236) & 3231(730) & 2556 & 1970 \\
25 X 50 & 0.98(0.64) & 0.76(0.79) & 3.12 & 4.00 & 0 & 1 & 5522(1851) & 4478(1968) & 7367 & 5983 \\
25 X 100 & 1.22(0.76) & 1.02(0.80) & 3.52 & 3.44 & 0 & 0 & 6704(2977) & 4824(3691) & 22793 & 15582 \\
\midrule
50 X 5 & 4.54(1.21) & 2.74(0.82) & 8.40 & 4.80 & 0 & 0 & 2805(825) & 3111(2346) & 652 & 644 \\
50 X 10 & 1.63(0.78) & 0.85(0.49) & 3.80 & 2.00 & 0 & 1 & 4818(1732) & 3410(1566) & 1746 & 1518 \\
50 X 25 & 1.72(0.66) & 0.90(0.37) & 3.36 & 1.60 & 0 & 0 & 7604(1662) & 5735(1714) & 7030 & 4764 \\
50 X 50 & 1.61(0.45) & 0.97(0.36) & 2.76 & 1.92 & 0 & 0 & 8573(1729) & 7110(2077) & 18569 & 13504 \\
50 X 100 & 1.50(0.60) & 1.01(0.44) & 2.96 & 2.72 & 0 & 0 & 9178(705) & 8132(1189) & 52250 & 39889 \\
\midrule
100 X 5 & 5.14(1.03) & 3.32(0.74) & 7.60 & 5.60 & 0 & 0 & 5442(1610) & 4417(976) & 2876 & 1843 \\
100 X 10 & 2.36(0.57) & 1.03(0.45) & 3.50 & 1.80 & 0 & 0 & 8590(1161) & 6651(1594) & 6909 & 4301 \\
100 X 25 & 2.52(0.58) & 1.19(0.42) & 4.24 & 2.16 & 0 & 0 & 9675(146) & 8769(3070) & 20390 & 12047 \\
100 X 50 & 2.51(0.59) & 1.23(0.34) & 4.18 & 2.28 & 0 & 0 & 9641(78) & 9319(387) & 46193 & 30645 \\
100 X 100 & 2.36(0.47) & 1.26(0.33) & 3.61 & 1.95 & 0 & 0 & 9708(245) & 9702(808) & 129312 & 87792 \\
\bottomrule
\end{tabularx}
\end{table*}

As it can be seen in Tables \ref{table:Alpha2}, \ref{table:Alpha15} the local search based on the use of subgraphs $GrowN$ manages to significantly outperform $GrowR$. As expected, $GrowN$ produces a higher level of improvement for problem instances having a higher number of roots, or in other words when the  graph is divided into a greater number of subgraphs. In case of problems having 100 subgraphs the  average error is nearly halved. For problem instances with lower values of $n$ this improvement is smaller but consistent. In the 50 problem sets, in only 5 $GrowR$ produces slightly better results. This only occurred for problems having five subgraphs, in which the use of neighborhoods is not essential. Overall both methods had a good performance, with an average error going up to $9\%$/$5\%$ and $5\%$/$3.5\%$ for problem instance generated using $\alpha = 2/1.5$ for $GrowR$ and $GrowN$, respectively. In practice the problem instances having the maximal number of allowed nodes $M = 5$ proved to be the hardest. The methods managed to find better solutions for  graphs with higher edge densities ($\alpha = 1.5$). We believe that the main reason for this is that such problem instances are in practice easier to solve since there is a greater number of potential bi-connected subgraphs.  The hardest instances to solve are the ones in which the maximal allowed size of a graph is the smallest and the number of subgraphs is the highest. For a notable number of test instances $GrowN$ manages to find optimal solutions, in case of $\alpha = 2/1.5$ it is close to $20\% / 30\%$. It is important to note that optimal solutions are only found for graphs having up to 500 nodes. $GrowN$ has a robust behavior in the sense that the maximal error for a problem set  is $9.6\% / 5.6 \%$ but in the majority of the cases it is less than $5\% /3\%$ for $\alpha = 2/1.5$.

When we observe the computational speed of the proposed methods the additional cost for calculating the neighborhoods in $GrowN$ is overall neglectable  when compared to $GrowR$. In case of large problem instances the use of neighborhoods even decreases the calculation time. Overall both methods prove to be computationally  very efficient having taken  100/120 and 80/87 seconds to generate close to 10 000 solutions in case of graphs having 10 000 nodes and $\alpha = 2/1.5$ for $GrowR$ and $GrowN$, respectively. It is important to note that for the largest instances in case of both methods the execution of the algorithm was terminated before the stagnation occurred (no improvements of the best found solution). The scaling of the methods is good in the sense that the increase of average execution time from instances having 100 nodes to 10000 was around 10 000 times. The number of iterations performed for different problem instances for the same pair $n,M$ would highly vary which can be seen from the standard variation. We wish to point out that  $GrowN$ manages to find high quality solutions with a  low number of generated solutions when compared to the solution space. 

\section{Conclusion}

In this paper we have introduced a new problem of finding the maximal number of nodes contained in a set of disjoint bi-connected subgraphs of a graph with the additional constraint on the maximal  size of a subgraph. This type of problem can be potentially applied for many practical problems. One example is the partitioning of electrical grids into a system of interconnected microgrids with a high level of resistance to failure. For solving the MBCPG-SC, a novel computationally efficient method for growing bi-connected subgraphs has been introduced. This method has been adapted to the setting of growing multiple graphs in parallel to generate solutions for MBCPG-SC. The quality of solutions generated in this way  was further improved using a local search method exploiting neighboring relations between subgraphs. The proposed method managed to  acquire  approximate solutions having an average error of up to $5\%$ when compared to known optimal solutions.  Further, the method is highly computationally efficient in the sense that it manages to find such solutions within two minutes for graphs having 10.000 nodes.

In the future we plan to extend the current research in several directions. First, we aim to  explore the weighted version of the problem which would be more suitable for problems occurring in electrical distribution systems. Once the weighted version is solved we may also relate the problem towards  multi-depot vehicle rooting problems and the assignment of costumers to routes of different depots. Secondly, we shall extend the method to metaheuristic approaches like ant colony optimization, GRASP or variable neighborhood search which appear  as good options. Finally, we shall explore the potential of applying the proposed growth procedure for problems like the  bi-connected dominating set and bi-connected  vertex cover problem.


\bibliographystyle{spbasic}      

%
%

\end{document}